\documentclass[preprint,11pt,authoryear]{elsarticle}

\usepackage[T1]{fontenc}
\usepackage[utf8]{inputenc}
\usepackage{lmodern}
\usepackage{amssymb}
\usepackage{amsmath}
\usepackage{booktabs}
\usepackage{tabularx}
\usepackage{longtable}
\usepackage{array}
\usepackage{multirow}
\usepackage{threeparttable}
\usepackage{enumitem}
\usepackage{xcolor}
\usepackage{graphicx}
\usepackage{microtype}
\usepackage[hidelinks]{hyperref}
\usepackage{float}
\usepackage[letterpaper,margin=1in]{geometry}

\newcolumntype{Y}{>{\raggedright\arraybackslash}X}

\begin{document}

\begin{frontmatter}

\title{Public trust in generative artificial intelligence in China: A mixed-methods scale development and validation study}

\author[aff1]{Zhiyuan Lai}
\author[aff1]{Wanxin Ran}
\author[aff1]{Pu Yan\corref{cor1}}
\ead{puyan@pku.edu.cn}
\author[aff2]{Zhouhan Chen}
\cortext[cor1]{Corresponding author.}
\affiliation[aff1]{organization={Department of Information Management, Peking University},
                   addressline={No. 5, Yiheyuan Road, Haidian District},
                   city={Beijing},
                   postcode={100871},
                   country={China}}
\affiliation[aff2]{organization={Safe Link Network},
                   city={New York},
                   country={United States}}

\begin{abstract}

Public engagement with generative artificial intelligence (GenAI) combines widespread adoption with cautious evaluation, yet existing measures differ in what they conceptualize as trust. This study develops and evaluates a measure of public trust in GenAI in China and situates individual trust judgments within broader public discourse. Using a multistage mixed-methods design, we refined a literature-derived item pool through a pilot survey ($n = 127$) and 68 semi-structured interviews, evaluated the scale in a survey ($N = 934$), and analyzed a corpus of 595,403 Weibo posts from 2023 to 2024. Psychometric analyses yielded a ten-item instrument comprising three correlated dimensions: instrumental, relational, and transparency-grounded trust. The trust composite was positively associated with AI literacy, frequency of AI use, and breadth of use. While the measure captures judgments concerning capability, relational responsiveness, and transparency, privacy, copyright, institutional accountability, and task stakes emerged as broader considerations surrounding trust and reliance in interviews. Weibo discourse showed a decline in attention to work and productivity and growing attention to education and emotional interaction, while source- and topic-standardized evaluative stance showed no systematic shift. This study advances the conceptualization and measurement of public trust in GenAI and situates individual trust judgments within the broader social and informational contexts of its use.

\end{abstract}


\end{frontmatter}

\section{Introduction}
\label{sec:introduction}

Generative artificial intelligence (GenAI) has rapidly entered everyday communication and information practices, supporting tasks from information seeking and content creation to learning and interpersonal support \citep{Baek2023Is,skjuve2024Why}. Widespread adoption, however, has been accompanied by uneven public responses. Cross-national surveys and social-media studies show that optimism about GenAI's productivity and creative utility coexists with concerns over misinformation, fairness, and accountability, with evaluations varying across contexts and user groups \citep{KelleyEtAl2021EightCountries,brauner2025Mapping,yang2025AI,chen2026See,Xu2024public,zhou2026AI,cohen2026ChatGPT}. In China, nationally representative evidence similarly documents a pattern of ``vigilant trust,'' in which favorable expectations toward AI sit alongside sustained awareness of potential risks \citep{DingEtAl2026VigilantTrust}.

Trust provides a useful lens for understanding these mixed evaluations. It concerns whether users regard a system as sufficiently dependable and appropriate to warrant confidence under uncertainty, which adoption or use alone cannot establish \citep{LeeSee2004TrustAutomation,McKnight2011SpecificTechnology,GliksonWoolley2020HumanTrustAI}. Users may routinely use a convenient chatbot while remaining cautious about its outputs, and their willingness to rely on it may vary with task stakes \citep{alon2026Trusting,zhang2026Crossdomain}. Understanding public trust in GenAI therefore requires examining the grounds on which people extend trust and the circumstances under which they rely on these systems.

Existing instruments leave this question unresolved. Scales developed for traditional automation and software emphasize competence, reliability, and helpfulness \citep{Jian2000TrustAutomation,McKnight2002TrustMeasures,McKnight2011SpecificTechnology,McGrath2025Measuring}, whereas GenAI introduces open-ended dialogue, probabilistic outputs, and limited visibility into evidential provenance \citep{Baek2023Is,JacoviEtAl2021Formalizing,kidd2023How}. Recent GenAI measures incorporate relational, epistemic, or transparency-related attributes, yet they differ in how they position these features relative to trust \citep{pandey2026Epistemic,rafikova2026Measuring,wang2026Validation}. Measuring public trust in GenAI therefore requires clarifying how core trust judgments are structured and how they differ from the broader risks, conditions, and reliance practices surrounding AI use.

Different empirical traditions capture distinct layers of this problem. Computational studies map socially visible concerns and evaluative patterns at scale \citep{Xu2024public,zhou2026AI,cohen2026ChatGPT}; qualitative interviews trace how users interpret and negotiate trust in everyday interaction \citep{skjuve2024Why}; and psychometric surveys test whether individual evaluations form a reproducible latent structure \citep{McGrath2025Measuring,pandey2026Epistemic,wang2026Validation}. The prominence of an issue in public discourse may differ from its significance in individual trust judgments. Integrating these forms of evidence can help clarify what a measure of GenAI trust captures and how it relates to broader public concerns.

To address these gaps, this study examines how to conceptualize and measure public trust in GenAI within its broader social context. Three related research questions guide this inquiry:

\textbf{RQ1.} How do people understand trust in GenAI, and what considerations inform their judgments about whether and when to trust it?

\textbf{RQ2.} How are trust-related judgments about GenAI structured at the individual level, and how can they be meaningfully measured?

\textbf{RQ3.} What concerns and evaluations characterize public discourse on GenAI, and how do they contextualize individual trust judgments?

This study addresses these questions through a multistage mixed-methods design in China. A literature-derived pool of 14 items was evaluated in a pilot survey ($n=127$) and refined through 68 semi-structured interviews. The revised items were evaluated through exploratory and confirmatory factor analyses using data from a two-wave survey ($N=934$). A retrospective computational analysis of 595,403 Weibo posts examined public discourse on the social impacts of GenAI between January 2023 and September 2024. This study contributes a ten-item measure of GenAI trust encompassing instrumental, relational, and transparency-grounded dimensions. It also demonstrates how evidence from situated experiences and public discourse can inform the development of a measurement instrument and clarify the scope of the construct it represents.

\section{Literature review and theoretical framework}
\label{sec:literature}

\subsection{Conceptualizing public trust in AI}
\label{subsec:concept}

Trust is commonly understood as a positive expectation that enables a person to accept vulnerability or uncertainty in relation to another party \citep{mayer1995Integrative,kramer1999TRUST,lewis1985Trust}. In interactions with automated and intelligent systems, such expectations concern whether the technology will behave dependably enough to justify confidence in its use \citep{LeeSee2004TrustAutomation,McKnight2011SpecificTechnology}. Trust in technology is related to reliance, while technology use may also reflect convenience, organizational mandates, switching constraints, or the absence of viable alternatives \citep{LeeSee2004TrustAutomation,brown2002Really,hargittaiWhata}. Nevertheless, trust provides an important basis for whether users are willing to consider, adopt, disclose information to, or continue interacting with an AI system \citep{AfrooghEtAl2024TrustAI,salih2025AIEnabled}.

Trust calibration is critical in human-AI interaction because both under-trust and over-trust carry costs (Lee and See, 2004). Insufficient trust can deter beneficial use, whereas excessive trust can lead users to accept inaccurate outputs, delegate unsuitable tasks, or disclose sensitive information without scrutiny (Ng and Zhang, 2025; Ferrario, 2026). In this study, public trust in GenAI refers to ordinary users' generalized positive expectations that GenAI can perform and interact in ways that warrant confidence under conditions of uncertainty. For generative systems, such expectations may rest on multiple grounds, including perceived capability and relational responsiveness as well as epistemic resources, such as explanations, traceable sources, and opportunities for verification. Standardized measurement makes it possible to examine how these grounds are organized and how trust varies across social groups, contexts, and patterns of AI use.

\subsection{An integrative framework of trust-relevant judgments in GenAI}
\label{subsec:distinctive}

The framework combines established trustworthiness models with GenAI's interactional and epistemic characteristics. Interpersonal and technology-trust research identifies competence, benevolence, and integrity as beliefs about whether a trustee is capable, oriented toward the trustor's interests, and guided by acceptable principles \citep{mayer1995Integrative,McKnight2011SpecificTechnology,lankton2015Technology}. Because GenAI communicates in open-ended language and produces probabilistic content, users may also evaluate social-emotional responsiveness and the transparency and verifiability of outputs \citep{hu2021Dual,deFreitasEtAl2023Psychological,JacoviEtAl2021Formalizing,li2025Warmth}. These five candidate domains informed the preliminary item pool.

Competence concerns whether GenAI can understand requests and produce relevant, effective, and sufficiently dependable responses across tasks. It extends established emphases on functionality and reliability to language understanding, personalization, and response adequacy \citep{LeeSee2004TrustAutomation,McKnight2011SpecificTechnology,Baek2023Is}. Empirical studies similarly associate perceived efficiency, personalization, functionality, and competence with trust and usage intentions
\citep{Baek2023Is,li2025Warmth,choung2023Trustb,huynh2025Generative}.

Benevolence concerns whether a system appears oriented toward users' interests and well-being, whereas integrity concerns whether it appears truthful and consistent with principles that users regard as acceptable. In conversational interaction, users may therefore judge whether GenAI tries to help and acts in their interests, as well as whether it respects facts and responds honestly \citep{mayer1995Integrative,lankton2015Technology,hu2021Dual,kramer2025Tricking,novozhilova2024More}.

Social-emotional responsiveness refers to the perceived ability of GenAI to recognize communicative and affective cues and respond in an attentive, appropriate, and socially intelligible manner. Through natural-language interaction, GenAI can create a sense of social presence and invite users to evaluate whether the system listens, understands, and responds appropriately \citep{hu2021Dual,GliksonWoolley2020HumanTrustAI,deFreitasEtAl2023Psychological}. It is related to anthropomorphism but does not require users to attribute consciousness or human identity to the system. Social presence and emotional engagement may support human-like trust, although anthropomorphic cues do not uniformly increase trust and can sometimes evoke discomfort \citep{Baek2023Is,huynh2025Generative}.

Transparency and verifiability concern the epistemic resources through which users can understand relevant aspects of system operation, interpret outputs, identify information sources, and check generated claims against external evidence \citep{JacoviEtAl2021Formalizing,shin2021Effects,scharowski2023Exploring,cetinkaya2026Transparency}. These resources are especially important for generative systems, whose fluent outputs may conceal uncertainty or unsupported claims \citep{kidd2023How}. Their relevance to trust depends on whether they enable meaningful scrutiny: accurate explanations and traceable sources may provide grounds for warranted confidence, whereas superficial explanations or unrelated citations may instead foster unwarranted trust \citep{scharowski2023Exploring,ding2025Citations,chen2026See}. Accordingly, the present study focuses on the extent to which users ground their trust in explanations, traceable sources, and opportunities for verification. This transparency-grounded trust captures the epistemic conditions under which confidence in GenAI is regarded as warranted.

\subsection{Methodological approaches to conceptualizing and measuring AI trust}
\label{subsec:existing}

Existing AI trust research draws on computational text analysis, qualitative inquiry, and survey measurement to examine trust across three analytical levels: visible public discourse, situated user experiences, and individual survey responses.

Computational studies identify the public contexts in which emerging technologies acquire social meaning by mapping issue salience, evaluative patterns, and discourse change at scale. Social-media and cross-channel analyses of GenAI have identified recurring concerns around utility, performance, misinformation, fairness, responsibility, and representation \citep{qi2024Excitements,wang2025Will,cohen2026ChatGPT,Xu2024public,naing2024Public,zhou2026AI}. Such evidence can reveal socially salient contexts that predefined instruments may overlook, but it captures discourse rather than individual trust.

Qualitative evidence locates trust within objects, tasks, and practices. Open-ended accounts from early ChatGPT users differentiated productivity, learning, creative, and social-support motives \citep{skjuve2024Why}. Interviews with active GenAI users further revealed tensions among practical utility, epistemic authority, verification, and platform control \citep{Brandtzaeg2026When}. AI-enabled health-app interviews similarly separated confidence in outputs from trust in companies and professional or regulatory endorsement \citep{RyanEtAl2025MHealth}. Such accounts support item meaning, wording, and construct boundaries.

Surveys test whether candidate content forms a reproducible individual-difference structure. Reviews nevertheless find heterogeneous definitions and warn against treating reliance as a direct trust measure \citep{VereschakEtAl2021Methods}. Recurrent operationalizations include competence and reliability, relational qualities, safety and risk, transparency and controllability, and global trust or use intentions \citep{Ng2025Trust}. Table~\ref{tab:existing-scales} summarizes key instruments identified as directly relevant to the conceptualization, operationalization, or differentiation of trust and adjacent constructs in AI and GenAI contexts.

\begingroup
\scriptsize
\setlength{\tabcolsep}{2.5pt}
\renewcommand{\arraystretch}{1.12}

\begin{longtable}{
  >{\raggedright\arraybackslash}p{2.30cm}
  >{\raggedright\arraybackslash}p{3.05cm}
  >{\raggedright\arraybackslash}p{5.10cm}}

\caption{Selected measures relevant to the conceptualization and measurement of trust in AI and GenAI}
\label{tab:existing-scales}\\

\toprule
Study & Focal construct and referent & Core dimensions or operationalization\\
\midrule
\endfirsthead

\multicolumn{3}{l}{\scriptsize\textit{Table \thetable\ continued}}\\
\toprule
Study & Focal construct and referent & Core dimensions or operationalization\\
\midrule
\endhead

\midrule
\multicolumn{3}{r}{\scriptsize Continued on next page}\\
\endfoot

\bottomrule
\multicolumn{3}{p{12.7cm}}{\scriptsize\textit{Note.} Panel A includes established measures used in the initial development of the item pool. Panel B includes subsequently published or contemporaneous measures used to situate the resulting instrument within the evolving GenAI-trust literature.}\\
\endlastfoot

\multicolumn{3}{l}{\textit{Panel A. Item-development sources}}\\
\addlinespace

\citet{McKnight2011SpecificTechnology}
& Trust in a specific technology
& Reliability, functionality, and helpfulness; trusting beliefs distinguished from propensity, structural assurance, situational normality, and use \\

\citet{lankton2015Technology}
& Trust in human-like and system-like technologies
& Human-like trust: competence, benevolence, and integrity; system-like trust: functionality, reliability, and helpfulness \\

\citet{hu2021Dual}
& Trust in conversational AI
& Competence, benevolence, and integrity \\

\addlinespace
\multicolumn{3}{l}{\textit{Panel B. Subsequent GenAI trust measures}}\\
\addlinespace

\citet{rafikova2026Measuring}
& Human--AI trust in a GenAI chatbot
& Understandability, competence, reliability, helpfulness, attachment, autonomy, faith, and institutional credibility \\

\citet{balazs2026Epistemic}
& Epistemic trust in GenAI
& Epistemic authority, epistemic reliability, and suspension of critical judgment \\

\citet{pandey2026Epistemic}
& Epistemic trust in GenAI in higher education
& Trustworthiness, contextual influences, dependability and safety, predictability and transparency, performance expectations, and user control \\

\citet{wang2026Validation}
& Trust in GenAI services
& Benevolence, competence, and reciprocity \\

\citet{brionesLlamoctanta2026CEIA}
& Emotional trust in GenAI
& Perceived authenticity, emotional validation, affective recognition, and symbolic bonding \\

\citet{huynh2025Generative}
& Human-like cognitive trust in workplace GenAI
& Competence, benevolence, and integrity \\

\citet{lee2025Dynamic}
& Overall trust in ChatGPT
& Global trust; functionality, helpfulness, predictability, and transparency modeled as distinct system attributes \\

\citet{McGrath2025Measuring}
& Trust in automated and AI systems
& Confidence, reliability, and global trust; intended reliance treated as a related but distinct outcome \\

\end{longtable}
\endgroup

Combining these three lines of evidence allows this study to specify core trust dimensions, test their psychometric structure, and situate them within public discourse.

\section{Method}
\label{sec:method}

\subsection{Research design and timeline}
\label{subsec:design}

\begin{figure}[H]
\centering
\includegraphics[width=0.92\textwidth]{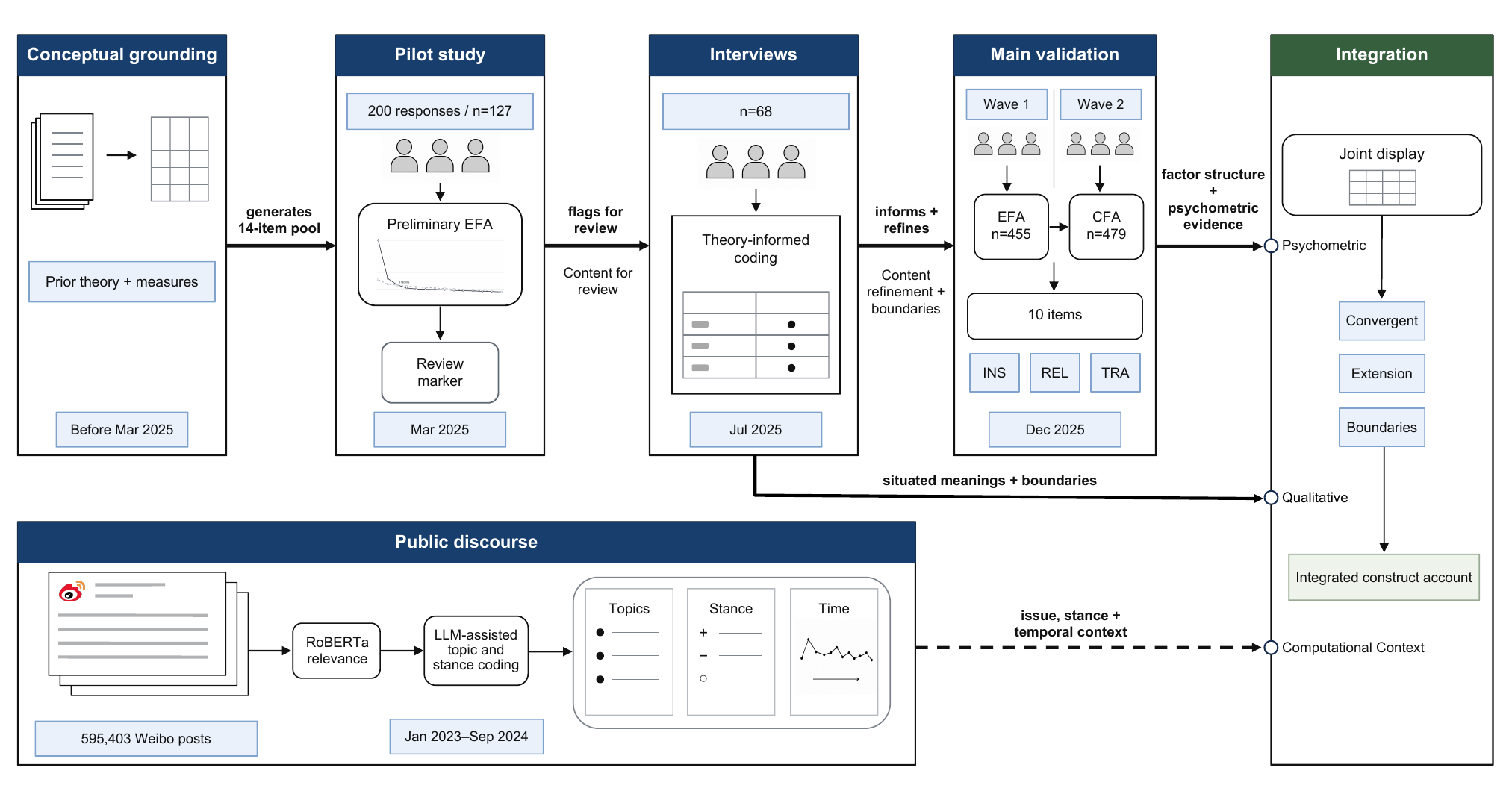}
\caption{Research timeline.}
\label{fig:timeline}
\end{figure}

This study employed a multistage mixed-methods design to develop and evaluate a measure of public trust in GenAI and to situate the resulting measure within Weibo discourse \citep{creswell2017designing,fetters2013Achieving,schoonenboom2017How}. The design integrated four components: a March 2025 pilot survey and July 2025 semi-structured interviews to refine item content, a December 2025 survey to evaluate factor structure and psychometric validity, and a retrospective computational analysis of Weibo posts to map the surrounding public discourse. The research timeline is shown in Figure~\ref{fig:timeline}.

The mixed-methods design served the purposes of development, complementarity, and expansion \citep{greene1989Conceptual}. Integration occurred through both building and merging: interview findings informed the review and refinement of candidate items before the main survey, while qualitative, psychometric, and computational findings were subsequently brought together through a joint display to compare areas of convergence, extension, and construct boundaries \citep{fetters2013Achieving}.

\subsection{Item development and qualitative content analysis}
\label{subsec:item-development}

\subsubsection{Initial item pool and pilot study}
\label{subsubsec:item-method}

The preliminary item pool was developed deductively from established trust theory and existing instruments \citep{McKnight2011SpecificTechnology,lankton2015Technology,hu2021Dual,deFreitasEtAl2023Psychological}. Fourteen candidate items represented five provisional content domains. Competence concerned capability, stable performance, and the fulfillment of user needs. Benevolence concerned helpful intent and attention to users' interests and well-being, whereas integrity concerned truthfulness and non-deception. Social-emotional responsiveness addressed warmth, emotional understanding, and socially appropriate communication. Transparency and verifiability concerned explanations of system operation and outputs and the provision of traceable information sources. These domains defined the content breadth of the initial pool rather than a prespecified factor structure. Items were rated on a five-point scale from 1 (strongly disagree) to 5 (strongly agree). 

The 14-item pilot questionnaire was administered in March 2025 through the Tencent Questionnaire platform, with participants recruited through its sample service. Of 200 submitted responses, 127 were retained after excluding responses failing the attention checks. Preliminary item structure was examined using EFA based on polychoric correlations, minimum-residual extraction, and oblimin rotation. Parallel analysis indicated a three-factor solution. The pilot factor patterns, together with subsequent interview-based content review, informed wording refinement and clarification of provisional construct boundaries. All 14 candidate items were administered in the main survey. Supplementary Appendix~A reports the pilot pattern matrix and parallel analysis (Table~A.1 and Figure~A.1) and the item wording and retention record (Table~A.2).

\subsubsection{Interviews and coding}
\label{subsubsec:interviews}

The qualitative material comprised 68 semi-structured interviews collected in July 2025 as part of a broader field experiment conducted at a high school. Participants were students and teachers involved in the field study. Interviews were conducted one-to-one and in person, lasted approximately 10 minutes, and covered participants' AI use and experiences alongside questions concerning trust in AI. All interviews were conducted in Chinese, audio-recorded with informed consent, and transcribed for analysis. For the present study, the full transcripts were reanalyzed to identify all material relevant to AI trust and its implications for the candidate item pool.

The transcripts were analyzed by one coder using a theory-informed qualitative coding framework and organized in a case-by-theme matrix \citep{miles1994Qualitative,hsieh2005Three}. Analysis proceeded through transcript familiarization, coding of trust objects and evaluative criteria, identification of task and risk contexts and conditions of reliance, and cross-case comparison of recurrent and contrasting patterns. Coded passages were also classified according to their role in the focal construct, distinguishing trust judgments from perceived trustworthiness, conditions of trust, reliance practices, risks, consequences, and governance concerns.

Across participants, recurrent and contrasting accounts were mapped onto candidate items via the case-by-theme matrix to assess content coverage, clarify construct boundaries, and refine item wording \citep{boateng2018Best}. Supplementary Appendix~B reports participant characteristics (Table~B.1), the interview guide (Section~B.2), and the coding framework (Table~B.2).

\subsection{Main survey and psychometric analysis}
\label{subsec:survey-method}

The main survey was administered through Tencent Questionnaire in two waves in December 2025. Participants were recruited through the platform's sample service, received compensation, and provided informed consent before participation. Of the 1,227 responses, 293 cases were excluded for failing attention checks, reporting no prior awareness of GenAI, being younger than 18, or providing logically inconsistent demographic information, leaving a final sample of 934 (76.12\% of submitted responses). Wave 1 was used for EFA (Sample A; $n=455$), and wave 2 was reserved for CFA and measurement validation (Sample B; $n=479$). The screening flow is reported in Supplementary Appendix~C (Table~C.1).

The sample covered 30 provincial-level administrative divisions. Mean age was 29.41 years ($SD=8.92$; median $=27$; range $=18$--$81$). Women accounted for 59.9\% of respondents, 69.5\% were employed full-time, and 66.0\% held at least a bachelor's degree. Respondents reported an average of 15.02 years of internet use ($SD=6.58$) and 13.08 years of mobile-internet use ($SD=5.18$). Supplementary Appendix~C (Table~C.2) reports further sample characteristics.

EFA used maximum-likelihood extraction with oblimin rotation. Factor retention was guided by parallel analysis, eigenvalues, interpretability, and item content. Item evaluation additionally considered primary loadings $\geq .40$, cross-loadings $< .30$, primary--secondary loading differences $\geq .20$, and communalities $\geq .40$. A minimum-residual EFA based on polychoric correlations was also estimated to assess the robustness of the factor solution to the ordinal response format.

CFA compared the correlated three-factor structure with a one-factor alternative in Sample B under both maximum likelihood (ML) and weighted least squares mean- and variance-adjusted estimation (WLSMV), with WLSMV treating the items as ordered categorical variables. Model fit was evaluated using $\chi^2$, CFI, TLI, RMSEA with its 90\% confidence interval, and SRMR; AIC and BIC were additionally used for ML model comparisons. Standardized loadings and factor correlations were examined across estimators. Internal consistency was assessed using coefficient alpha and McDonald's omega. Composite reliability (CR) and average variance extracted (AVE) were used to assess construct reliability and convergent validity, while the Fornell--Larcker criterion and HTMT were used to assess discriminant validity. Supplementary Appendix~C (Section~C.4) reports the item loadings and latent factor correlations under both estimators.

Associations with external variables were examined in the full sample using separate cross-sectional regression models. The overall GenAI trust composite, calculated as the mean of the three subscale scores (INS, REL, and TRA), was related to AI-use frequency, AI-use breadth, and AI literacy. All models used standardized focal variables and adjusted for age, gender, education, employment, urban--rural residence, years of internet use, and years of mobile-internet use. The three lowest education categories were combined because of sparse observations. HC1 robust standard errors were reported. Supplementary analyses examined the three subscales separately and simultaneously and used the unweighted mean of all ten items as an alternative composite. Supplementary Appendix~D reports the external variables (Section~D.1), scoring rules (Section~D.2), and supplementary regressions (Section~D.3).

\subsection{Retrospective computational mapping of public discourse}
\label{subsec:computational-method}

\subsubsection{Corpus construction and relevance classification}

The computational analysis used public Weibo posts published between 1 January 2023 and 1 September 2024 and retrieved with 18 GenAI-related search terms. Exact- and near-duplicate removal yielded 595,403 unique posts. Posts were classified as relevant when they substantively discussed the consequences of AI for social groups, institutions, everyday practices, or public outcomes. Product announcements, technical specifications, tutorials, marketing, financial information, and entertainment content without an identifiable social consequence were excluded. Supplementary Appendix~E (Section~E.1) lists the retrieval terms and documents deduplication.

Two coders independently annotated an initial set of 700 posts, with Cohen's $\kappa$ and nominal Krippendorff's $\alpha$ both equal to .693. Disagreements were adjudicated, and six active-learning rounds expanded the labeled set to 2,200 posts using a combination of uncertainty sampling, high-confidence positive sampling, and random sampling. The final classifier fine-tuned the Chinese RoBERTa-wwm-ext checkpoint (\texttt{hfl/chinese-roberta-wwm-ext}) \citep{liu2019RoBERTa,cui2021PreTraining} using 1,632 training, 288 validation, and 280 report-test cases. Class weighting addressed class imbalance, and validation probabilities were calibrated using temperature scaling \citep{guo2017Calibration}. The classification threshold was set at .425 before evaluation on the report-test set. The training and evaluation details are reported in Supplementary Appendices~E (Section~E.2) and~F (Section~F.1).

\subsubsection{LLM-assisted topic and stance coding}

The RoBERTa classifier identified 203,657 candidate-relevant posts, which were subsequently coded through a custom GABRIEL-style workflow using deepseek-v4-flash with temperature set to 0 \citep{AsirvathamEtAl2026Gabriel}. A topic codebook was developed iteratively through human review of sampled posts and refinement of topic definitions and boundary rules before full-corpus coding \citep{hsieh2005Three}. The final codebook contained 12 non-exclusive substantive topics and a thirteenth boundary category. For each post, the model assigned one or more topic labels, a primary topic, a stance category, a confidence score, and a review flag. The coding pipeline and raw outputs were versioned for reproducibility. The codebook and decision rules are documented in Supplementary Appendix~E (Section~E.3).

Coding performance was evaluated against a stratified human-coded sample of 200 posts. Evaluation metrics included exact match, Jaccard similarity, micro- and macro-F1 for multi-label topics, and accuracy and Cohen's $\kappa$ for primary-topic and stance classifications. Substantive topic analyses used 152,485 posts classified by the LLM workflow as social-impact-related and assigned a non-boundary primary topic; stance analyses used the 151,083 posts that additionally received a negative, neutral, or positive stance. Multi-label topic proportions were calculated independently for each topic, whereas primary topics were mutually exclusive. The validation sample, construction of the final human reference, and detailed agreement results are reported in Supplementary Appendix~F (Sections~F.2--F.3).

\subsubsection{Topic structure, stance, and temporal analysis}

Topic-stance associations were examined separately from the multi-label descriptive summaries. For descriptive visualization, stance proportions were calculated within each of the 12 multi-label substantive topics; because a post could receive multiple topic labels, these topic-specific denominators were not mutually exclusive. Inferential analyses instead relied on each post's mutually exclusive primary-topic assignment. Among posts assigned to one of the 12 substantive primary topics and with a valid stance classification, we constructed a $12 \times 3$ contingency table and evaluated the association between primary topic and stance using a Pearson chi-square test of independence. Association magnitude was summarized using bias-corrected Cramér's $V$.

Temporal analyses were conducted on repeated cross-sectional posts from the 20 complete months between January 2023 and August 2024 (n=152{,}041 for topic analyses; 609 calendar days). For each of the 12 multi-label topics, a binary logistic regression modeled topic presence as a function of a linear month index (0= January 2023 to 19= August 2024) and categorical retrieval-keyword fixed effects. Inference used finite-sample-corrected standard errors clustered by calendar day, allowing arbitrary dependence among posts observed on the same day, and the 12 month-effect \(p\) values were adjusted using the Benjamini--Hochberg procedure. Odds ratios, average marginal effects, and source-standardized prevalence estimates were used to summarize effect magnitude.

Sensitivity analyses examined two-way clustering by calendar day and user-nickname proxy and daily-score HAC covariance with Bartlett kernels over 3-, 7-, 14-, and 30-day lags. Stable account identifiers were unavailable, so nickname-based clustering was treated as an account-proxy sensitivity. A concentrated surge in late-August 2024 deepfake-related discussion substantially altered the retrieval-source composition of that month. Event sensitivity therefore was assessed by excluding August 2024, August 29--31, posts retrieved through the \textit{deepfake} keyword, and observations after June 2024. All specifications re-estimated the 12 topic models and applied BH correction within specification. Categorical-month models provided source-standardized monthly estimates without imposing a linear trajectory. A complementary multinomial model of mutually exclusive primary topics tested the joint temporal change in overall issue composition using the same retrieval-keyword adjustment and dependence-aware covariance specifications.

Temporal change in stance was examined using multinomial logistic regression with neutral stance as the reference category and month, retrieval-keyword fixed effects, and primary topic as predictors. Calendar-day-clustered covariance was used for the main uncertainty estimates, with two-way clustering and HAC(7) examined as sensitivity specifications. Adjusted stance proportions were standardized to the pooled joint distribution of retrieval keywords and primary topics. Supplementary Appendix~G reports the model specifications and sensitivity checks.

\section{Findings}
\label{sec:findings}

\subsection{Interview findings and content implications}
\label{subsec:interview-results}

The interview analysis showed that trust in GenAI was rarely expressed as an undifferentiated judgment. Participants evaluated AI in relation to what it was expected to do, the consequences of error, and the resources available for examining its outputs. Capability and accuracy were recurring grounds for confidence, but their practical significance varied across tasks, while labeling, source disclosure, and human verification frequently entered participants' accounts as conditions for deciding whether confidence was warranted. Relational considerations also appeared in accounts of emotional support and human-like communication, although they were more salient in particular interactional contexts than in general discussions of factual reliability. Overall, the interviews clarified how trust judgments operate in everyday use and separated core trust evaluations from task conditions, reliance practices, and governance concerns.

A particularly consistent pattern concerned epistemic scrutiny. Participants described checking AI-generated information against other sources, looking for identifiable evidence, relying on human review, or lowering their reliance when outputs could not be independently assessed. P019, for example, linked confidence in consequential information to the availability of reference sources that could subsequently be checked, while P043 regarded human review as necessary because AI systems could produce errors without recognizing them. Such practices suggest that explanations and traceable evidence matter not simply because they make a system appear transparent, but because they provide resources for judging whether confidence in a particular output is warranted. However, this pattern was task-dependent. Some participants expressed relatively broad confidence in routine learning or low-stakes uses without specifying additional safeguards, and P025 noted that explicit AI identification could support trust in some settings while making personal interaction feel less natural in others. The interview evidence therefore supported framing transparency and verifiability as epistemic grounds on which trust may be conditioned. In item refinement, this evidence reinforced the shift toward conditional formulations linking explanations and traceable sources explicitly to respondents' trust.

Trust thresholds also varied across task and perceived consequence. Across cases, participants were generally more permissive when AI was used for exploratory learning, drafting, entertainment, or other readily reversible activities, but more cautious when outputs concerned news, public facts, or specialized knowledge. P051 drew this distinction directly: fictional stories involved few concerns about factual accuracy, whereas AI-generated news about public affairs elicited substantially greater skepticism. Similar contrasts appeared elsewhere. P102 expressed confidence in AI for practical work-related assistance while distrusting AI-generated news and social-media interactions, and P106 described using AI-generated drafts for relatively inconsequential work while checking specialized answers against academic sources when accuracy mattered. Task stakes and perceived risks thus adjust the threshold at which users act on trust, shaping how trust is expressed without forming separate dimensions of the construct. They also help explain why reliance practices varied even among participants who described themselves as broadly trusting: some used AI directly for low-stakes assistance but verified, restricted, or sought oversight for more consequential tasks.

The interviews further clarified the boundary between trust in GenAI and concerns directed toward the wider sociotechnical environment. Privacy, copyright, platform responsibility, and accountability were salient, but these concerns targeted data practices, developers, platforms, or governance arrangements instead of the system's immediate capability or interactional qualities. P044, for example, attributed responsibility for harmful outputs to human and platform oversight. At the same time, everyday trust objects were not always sharply separated. P030 expressed general distrust of GenAI while grounding that judgment in the unauthorized use of artists' work for model training, illustrating how institutional or data-governance concerns can spill over into broader system-level evaluations. Such cases illustrate how institutional concerns can inform trust in the system itself. Accordingly, privacy and accountability concerns were retained as trust-relevant contexts or safeguards, whereas capability, relational responsiveness, and epistemic scrutiny remained the principal candidate bases represented in the item pool.

In sum, capability, relational responsiveness, and epistemic scrutiny emerged as core bases of trust evaluation, whereas task stakes, privacy, copyright, and institutional accountability operated as surrounding risks and conditions. These distinctions guided item revision prior to the main survey.

\subsection{Item screening and exploratory factor structure}
\label{subsec:item-efa-results}

The formal survey administered 14 trust-related items. Sample A ($n=455$) was suitable for factor analysis (KMO=.944; Bartlett's $\chi^2(91)=3422.019$, $p<.001$), and parallel analysis supported a three-factor solution. Maximum-likelihood EFA with oblimin rotation explained 57.01\% of the variance. As shown in Table~\ref{tab:efa-initial-loadings}, the three competence items loaded primarily on one factor, the three transparency items on another, and Benevolence 3 clustered with the three social-emotional responsiveness items. In contrast, Benevolence 1 and 2 loaded onto the instrumental factor alongside the competence items, and the two integrity items failed to form a separate factor, likewise loading primarily on the instrumental factor; Integrity 2 additionally showed weak separation from the relational factor (loading gap=.127). Considering the loading pattern, item content, and factor interpretability, Benevolence 1, Benevolence 2, Integrity 1, and Integrity 2 were removed.

\begin{table}[H]
\centering
\caption{Exploratory factor pattern matrix for the initial 14-item pool in Sample A ($N=455$)}
\label{tab:efa-initial-loadings}
\small
\begin{tabular}{lrrrr}
\toprule
Item & Factor 1 & Factor 2 & Factor 3 & $h^2$ \\
\midrule
Competence 1 & \textbf{.619} & .164 & $-.029$ & .523 \\
Competence 2 & \textbf{.715} & .012 & .027 & .553 \\
Competence 3 & \textbf{.733} & $-.106$ & .102 & .545 \\
Benevolence 1 & \textbf{.594} & $-.003$ & .181 & .539 \\
Benevolence 2 & \textbf{.504} & .077 & .168 & .480 \\
Benevolence 3 & .297 & \textbf{.577} & $-.080$ & .585 \\
Integrity 1 & \textbf{.585} & .213 & $-.011$ & .552 \\
Integrity 2 & \textbf{.424} & .297 & .034 & .479 \\
Social-emotional responsiveness 1 & .130 & \textbf{.698} & $-.007$ & .625 \\
Social-emotional responsiveness 2 & .107 & \textbf{.616} & .122 & .601 \\
Social-emotional responsiveness 3 & $-.094$ & \textbf{.844} & .104 & .703 \\
Transparency 1 & .016 & .082 & \textbf{.710} & .594 \\
Transparency 2 & $-.000$ & .076 & \textbf{.758} & .643 \\
Transparency 3 & .103 & $-.074$ & \textbf{.713} & .561 \\
\bottomrule
\end{tabular}
\begin{minipage}{0.94\linewidth}
\footnotesize \textit{Note.} Maximum-likelihood extraction with oblimin rotation.
\end{minipage}
\end{table}

The EFA was then re-estimated on the retained ten items in the same sample. The refitted solution remained suitable for factor analysis (KMO=.912; Bartlett's $\chi^2(45)=2242.674$, $p<.001$) and explained 60.16\% of the variance. The three factors were interpreted as instrumental trust (INS), relational trust (REL), and transparency-grounded trust (TRA); the complete retained-item ML pattern matrix is reported in Supplementary Appendix~C (Table~C.3). Nine items met all prespecified ML-EFA criteria. Competence 3 showed the weakest separation, loading .424 on INS and .251 on TRA (difference=.174); in the polychoric/minres EFA, the corresponding loadings were .477 and .257 (difference=.220), satisfying the .20 criterion. Given this robustness result and the item's contribution to competence coverage, it was retained. The resulting ten-item, three-factor structure was fixed before CFA.

\subsection{Confirmatory factor structure, reliability, and retained scale}
\label{subsec:psychometric-results}

\begin{figure*}[t]
\centering
\includegraphics[width=0.88\textwidth]{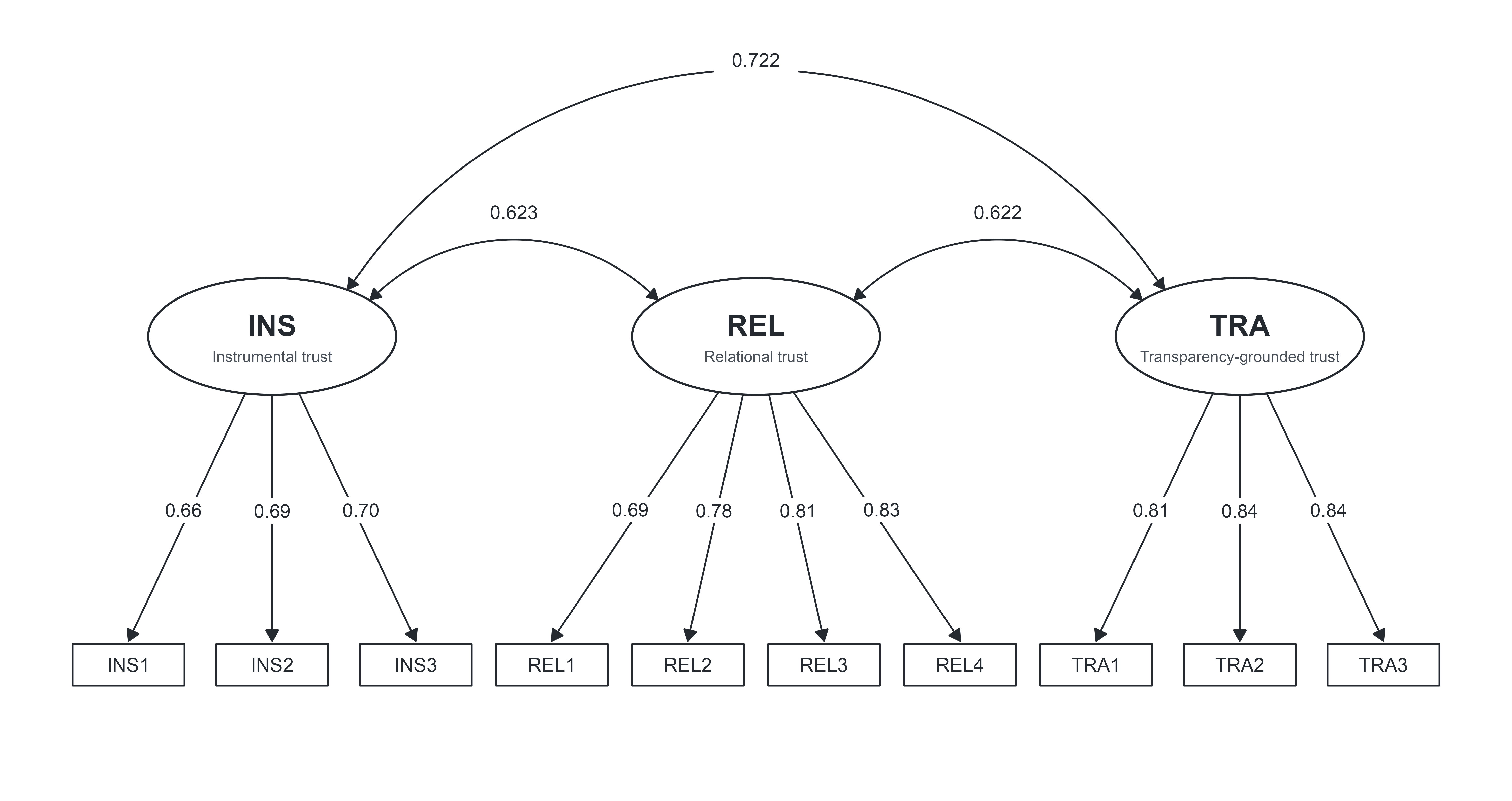}
\caption{Correlated three-factor confirmatory model in Sample B ($n=479$). Values on single-headed paths are standardized ML factor loadings, and curved double-headed paths are latent factor correlations. All factor loadings were significant at $p<.001$. INS = instrumental trust; REL = relational trust; TRA = transparency-grounded trust.}
\label{fig:cfa-model}
\end{figure*}

The correlated three-factor model fit substantially better than the one-factor model under both estimators (Table~\ref{tab:cfa-fit}). Under ML, the three-factor model showed acceptable fit and markedly lower AIC and BIC. Under WLSMV, its CFI, TLI, and SRMR were favorable, although RMSEA=.082 (90\% CI [.068, .096]) remained borderline. Standardized loadings ranged from .657 to .845 under ML and from .698 to .896 under WLSMV (all $p<.001$, Figure~\ref{fig:cfa-model}).

\begin{table}[H]
\centering
\caption{Comparison of confirmatory measurement models in Sample B ($n=479$)}
\label{tab:cfa-fit}
\small
\setlength{\tabcolsep}{3.5pt}
\resizebox{\textwidth}{!}{%
\begin{tabular}{llrrrrrrr}
\toprule
Estimator & Model & $\chi^2(df)$ & CFI & TLI & RMSEA (90\% CI) & SRMR & AIC & BIC \\
\midrule
ML & Correlated three-factor & 108.677 (32) & .967 & .953 & .071 [.056, .086] & .040 & 10484.944 & 10580.893 \\
ML & One-factor & 573.943 (35) & .767 & .701 & .179 [.167, .192] & .092 & 10944.210 & 11027.644 \\
WLSMV & Correlated three-factor & 134.122 (32) & .985 & .978 & .082 [.068, .096] & .040 & --- & --- \\
WLSMV & One-factor & 819.546 (35) & .881 & .847 & .217 [.204, .230] & .102 & --- & --- \\
\bottomrule
\end{tabular}
}
\begin{minipage}{0.98\linewidth}\footnotesize \textit{Note.} $N=479$. WLSMV entries are scaled fit statistics. All $\chi^2$ tests were significant at $p<.001$. AIC and BIC are not available for WLSMV.
\end{minipage}
\end{table}

Subscale reliability was acceptable to high, with alpha ranging from .727 to .869 and omega from .728 to .870 (Table~\ref{tab:reliability}). Composite reliability ranged from .725 to .870 under ML and from .781 to .912 under WLSMV. REL and TRA exceeded the conventional .50 AVE criterion under both estimators, whereas INS was borderline under ML (AVE=.468) but met the criterion under WLSMV (AVE=.543). Fornell--Larcker was not met for INS--TRA under either estimator, whereas all HTMT values were below .85 (range=.619--.708), indicating comparatively greater overlap between INS and TRA while supporting their empirical distinction.

\begin{table}[H]
\centering
\caption{Reliability, convergent validity, and factor distinctiveness}
\label{tab:reliability}
\small
\begin{tabular}{lrrrrrrr}
\toprule
Construct & Items & $\alpha$ & $\omega$ & CR$_{\mathrm{ML}}$ & AVE$_{\mathrm{ML}}$ & CR$_{\mathrm{WLSMV}}$ & AVE$_{\mathrm{WLSMV}}$ \\
\midrule
INS & 3 & .727 & .728 & .725 & .468 & .781 & .543 \\
REL & 4 & .858 & .861 & .861 & .609 & .890 & .671 \\
TRA & 3 & .869 & .870 & .870 & .690 & .912 & .776 \\
\bottomrule
\end{tabular}
\begin{minipage}{0.98\linewidth}\footnotesize \textit{Note.} CR = composite reliability; AVE = average variance extracted. HTMT: INS--REL=.619, INS--TRA=.708, and REL--TRA=.625.
\end{minipage}
\end{table}

The retained ten-item instrument represents three correlated bases of GenAI trust (Table~\ref{tab:items}). Instrumental trust captures confidence grounded in perceived capability, stable performance, and responsiveness to user needs. Relational trust captures confidence grounded in perceived care, emotional understanding, warmth, and socially appropriate communication. Transparency-grounded trust captures the extent to which confidence depends on explanations of system operation and outputs and access to traceable information sources. Each dimension can be scored as the mean of its constituent items. When an integrated measure of GenAI trust is required, an overall multidimensional composite can be calculated as the mean of the three subscale scores, giving equal weight to the three theoretically specified bases of trust.

\begin{table}[H]
\centering
\caption{English translation of the retained 10-item measure}
\label{tab:items}
\small
\renewcommand{\arraystretch}{1.15}

\begin{tabularx}{\textwidth}{p{2.9cm}Y}
\toprule
Dimension & Item \\
\midrule

\multirow[c]{3}{2.9cm}{\centering Instrumental trust}
& The performance of generative AI tools is stable and consistent when they interact with me. \\

& Generative AI tools can effectively understand my needs and provide responses that meet those needs. \\

& Generative AI tools have powerful capabilities. \\[0.6em]

\multirow[c]{4}{2.9cm}{\centering Relational trust}
& Generative AI tools care about my well-being and health. \\

& Generative AI tools can perform tasks requiring emotional understanding, such as counseling, emotional advice, or comfort. \\

& Generative AI tools have social competence and can communicate with people appropriately. \\

& Generative AI tools can show care and warmth toward me. \\[0.6em]

\multirow[c]{3}{2.9cm}{\centering Transparency-grounded trust}
& I would trust generative AI tools more if they could clearly explain how they work. \\

& I would trust generative AI tools more if they could explain their outputs. \\

& I would trust generative AI tools more if they could provide sources for the information they offer. \\

\bottomrule
\end{tabularx}

\begin{minipage}{0.98\textwidth}
\footnotesize
\textit{Note.} Items were rated on a five-point scale from 1 (strongly disagree) to 5 (strongly agree). The English items are translations of the Chinese-language instrument administered in the survey.
\end{minipage}
\end{table}

\subsection{Associations with external variables}
\label{subsec:nomological}

After covariate adjustment, the overall GenAI trust composite was positively associated with AI-use frequency ($\beta=.308$), AI-use breadth ($\beta=.178$), and self-reported AI literacy ($\beta=.690$; all $p<.001$; Table~\ref{tab:nomological}). The subscales exhibited different patterns of association. When entered jointly, TRA was negatively associated with AI-use frequency and breadth but remained positively associated with self-reported AI literacy (Supplementary Table~D.3). Together with the three-factor structure, these external associations support interpreting the 10-item scale as an integrated multidimensional measure capturing distinct but connected bases of GenAI trust.

\begin{table}[H]
\centering
\caption{Adjusted associations of the equal-weight multidimensional GenAI trust composite with external variables ($N=934$)}
\label{tab:nomological}
\small
\begin{tabular}{lrrrrr}
\toprule
Outcome & $\beta$ & Robust SE & 95\% CI & $R^2$ & $\Delta R^2$ \\
\midrule
AI-use frequency & .308 & .033 & [.243, .373] & .141 & .089 \\
AI-use breadth & .178 & .033 & [.113, .243] & .068 & .030 \\
AI literacy & .690 & .028 & [.635, .746] & .525 & .446 \\
\bottomrule
\end{tabular}
\begin{minipage}{0.96\linewidth}\footnotesize
\textit{Note.} The overall composite was calculated as the equal-weight mean of the INS, REL, and TRA subscale scores. Standardized coefficients are from separate cross-sectional models controlling for age, gender, education, employment, urban--rural residence, years of internet use, and years of mobile-internet use. HC1 robust standard errors are reported. $\Delta R^2$ indicates the increase in explained variance relative to the covariate-only model. All $p<.001$.
\end{minipage}
\end{table}

\subsection{Retrospective mapping of the public-discourse context}
\label{subsec:computational-results}

\begin{figure}[H]
\centering
\includegraphics[width=0.92\textwidth]{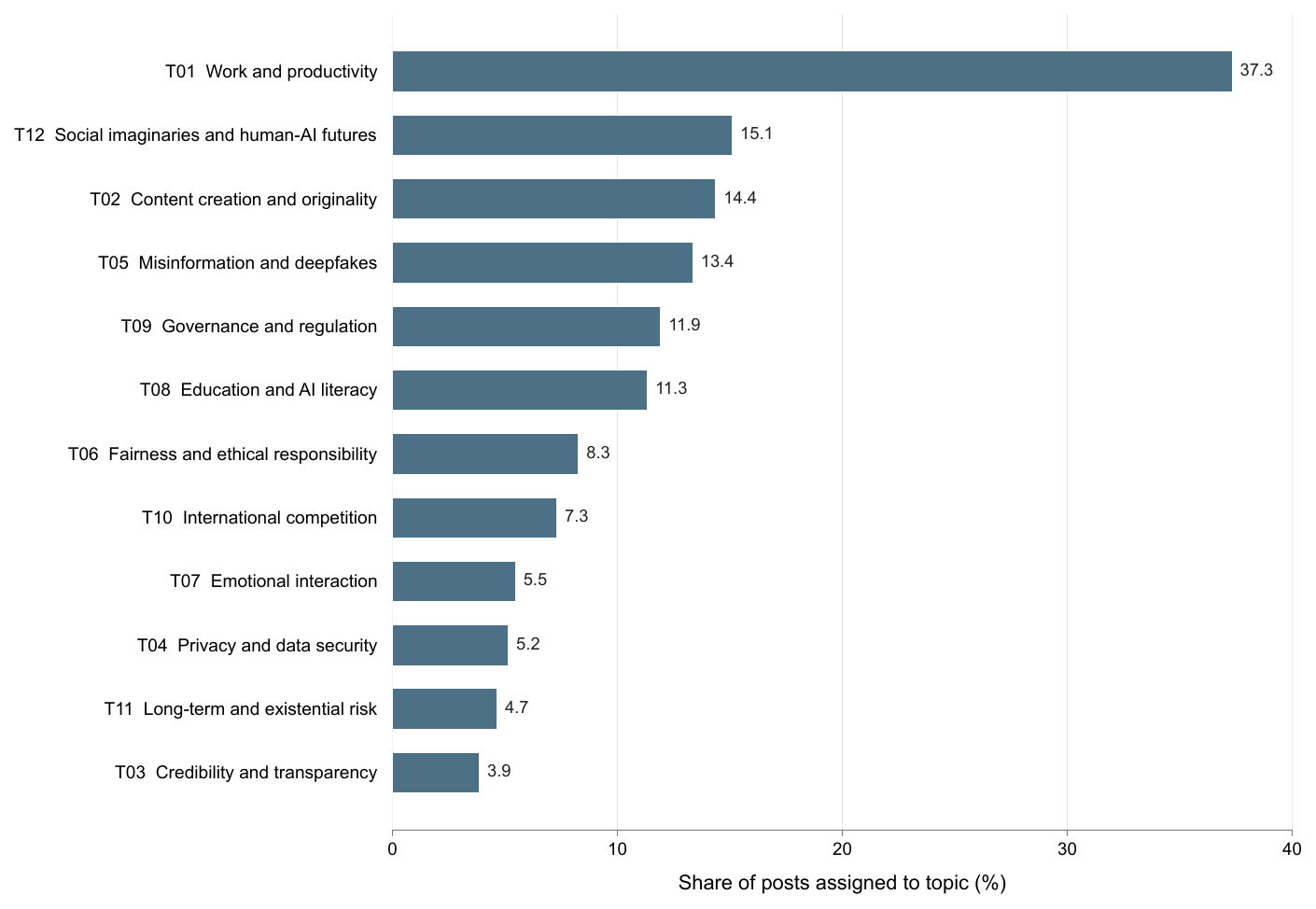}
\caption{Prevalence of substantive topics in the LLM-classified analytic corpus. Percentages are multi-label mention rates and therefore sum to more than 100\%.}
\label{fig:topic-distribution}
\end{figure}

The RoBERTa classifier identified 203,657 candidate social-impact posts from the 595,403-post corpus, with report-test accuracy=.843, precision=.710, recall=.795, F1=.750, and ROC--AUC=.918. LLM-assisted coding subsequently classified 152,485 posts as social-impact-related and assigned them a substantive primary topic; 151,083 also received a valid stance label. Human validation showed strong agreement for primary-topic coding (accuracy=.880, $\kappa=.869$) and stance coding (accuracy=.800, $\kappa=.716$), while multi-label topic coding achieved micro-F1=.750 and macro-F1=.771. Within the LLM-classified analytic corpus, work, productivity, and industrial application was the most prevalent topic (37.33\%). Substantial discussion also concerned social imaginaries and human--AI futures (15.11\%), content creation and originality (14.37\%), misinformation and deepfakes (13.35\%), governance and regulation (11.92\%), and education and AI literacy (11.33\%) (Figure~\ref{fig:topic-distribution}). Nearly one-third of posts (31.37\%) crossed topical boundaries. The strongest overlaps linked misinformation with fairness and ethical responsibility (Jaccard=.212; lift=3.43) and with privacy and data security (Jaccard=.140; lift=3.30), situating information integrity within a broader cluster of ethical, data-governance, and institutional concerns.

\begin{figure}[H]
\centering
\includegraphics[width=0.92\textwidth]{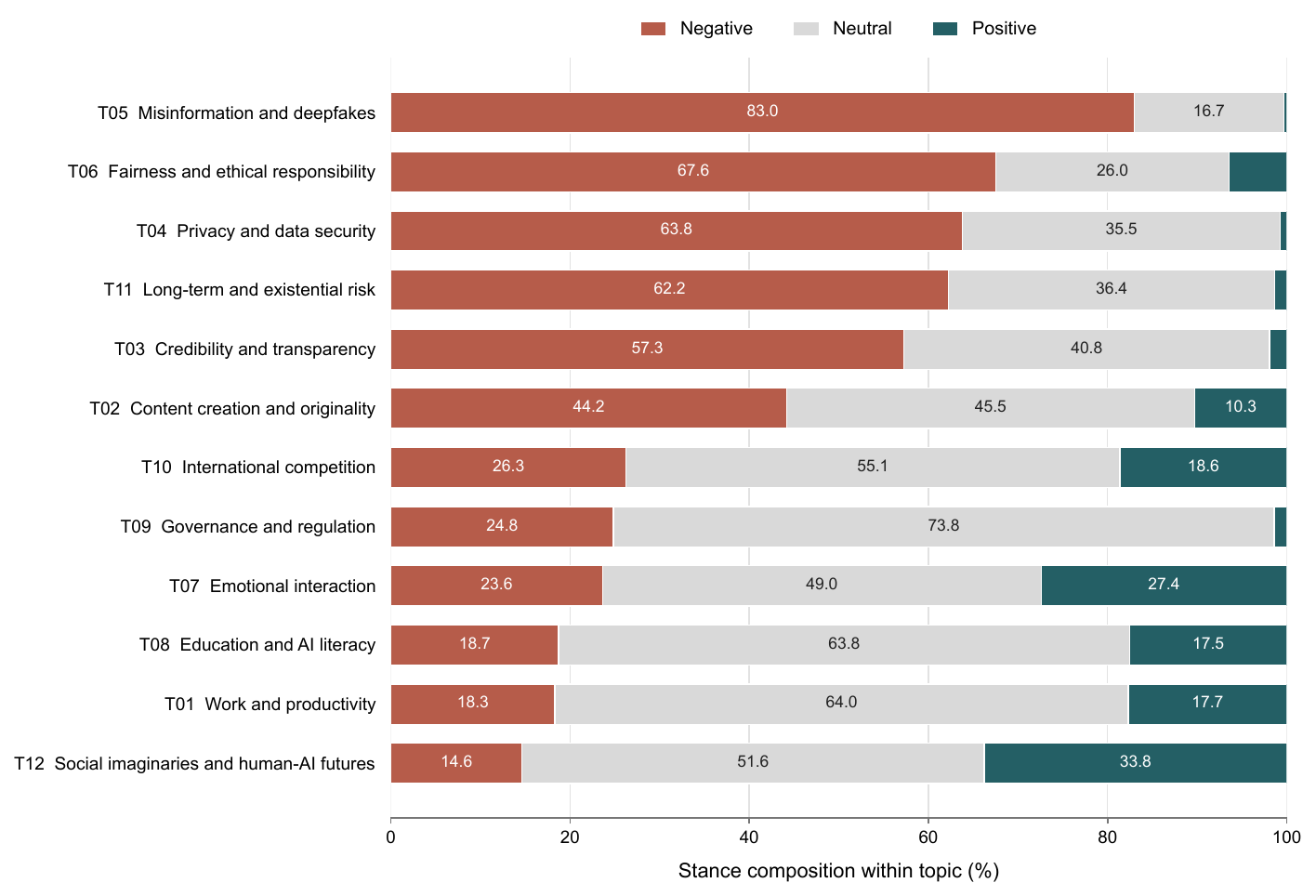}
\caption{Stance composition across substantive topics in the LLM-classified analytic corpus. Percentages are calculated separately within each multi-label topic among posts with a valid stance classification.}
\label{fig:topic-stance}
\end{figure}

The descriptive multi-label distributions in Figure~\ref{fig:topic-stance} showed substantial variation in stance composition across topics. In the omnibus test based on mutually exclusive primary-topic assignments, stance differed across topics ($\chi^2(22)=51{,}265.9$, $p<.001$, Cramer's $V=.412$). Risk-centered topics were predominantly negative: negative stances accounted for 83.0\% of misinformation and deepfake posts, 67.6\% of fairness and ethical-responsibility posts, 63.8\% of privacy and data-security posts, 62.2\% of long-term and existential-risk posts, and 57.3\% of credibility and transparency posts. By contrast, social imaginaries and human--AI futures had the largest positive share (33.8\%), followed by emotional interaction (27.4\%), whereas work and productivity, education and AI literacy, and governance and regulation were predominantly neutral or informational.

\begin{figure}[H]
\centering
\includegraphics[width=\textwidth]{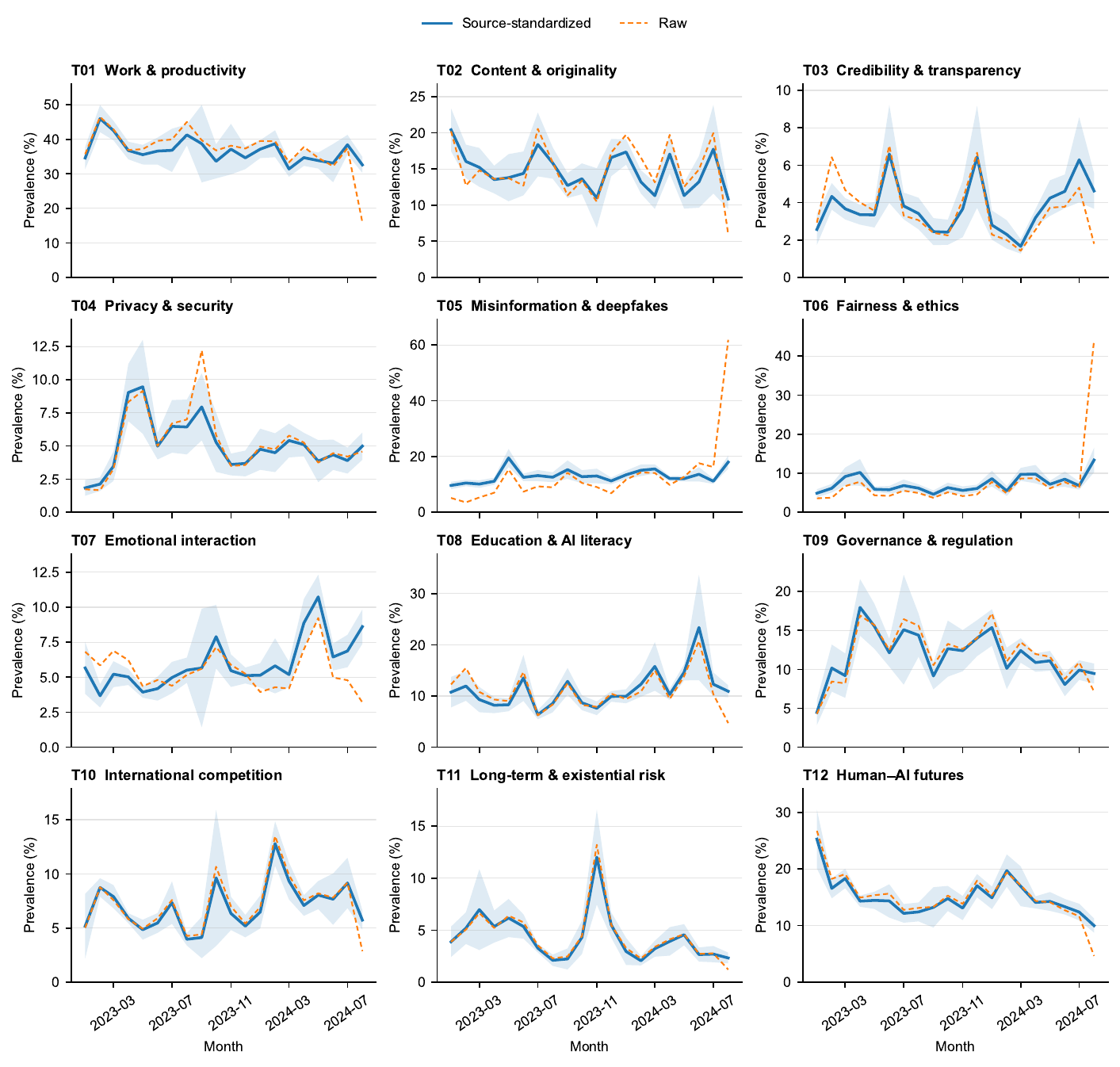}
\caption{Monthly prevalence of the 12 substantive topics, January 2023--August 2024. Solid lines show source-standardized estimates, dashed lines show raw prevalence, and shading denotes 95\% confidence intervals.}
\label{fig:topic-trends}
\end{figure}

The composition of observed discourse changed over time after adjustment for retrieval-source composition (Figure~\ref{fig:topic-trends}). A multinomial model of mutually exclusive primary topics showed a significant joint month effect under calendar-day clustering, two-way day--nickname clustering, and HAC(7) inference (all $p<.001$). In the multi-label binary topic models, the largest source-standardized changes were a decline in work and productivity (8.32 percentage points across the observation period) and increases in education and AI literacy (+6.56 points) and emotional interaction (+3.97 points); these linear topic effects remained significant after BH adjustment across all reported HAC bandwidths. These results provide the clearest evidence of sustained changes in issue composition.

Trends for several topics were less stable across dependence specifications. Misinformation and deepfakes showed a positive estimate (AME =+0.111 percentage points per month; day-cluster $q=.025$), and social imaginaries and human--AI futures showed a negative estimate (AME $=-0.149$; $q=.008$); however, neither remained significant under the daily-score HAC specifications. Fairness and ethical responsibility also had a positive point estimate (AME =+0.089) but did not pass BH adjustment under calendar-day clustering ($q=.107$). Event-window analyses preserved the directions of these estimates but showed that their inferential strength, particularly for fairness, depended on late-period event concentration and the observation endpoint (Supplementary Appendix~G, Tables~G.2--G.3). Supplementary Table~S1 reports the full set of topic-by-specification estimates.

Temporal variation was clearer in issue composition than in standardized stance. After standardization to a common retrieval-source and primary-topic composition, the predicted negative share increased by 0.32 percentage points across the observation period (95\% CI $[-1.82,2.47]$) and the positive share by 1.49 points (95\% CI $[-.79,3.77]$). The joint month effect was not significant under calendar-day-clustered inference ($p=.159$), two-way day--nickname clustering ($p=.190$), or HAC(7) inference ($p=.340$). The analysis therefore provided no evidence of a systematic temporal shift in stance after accounting for issue and retrieval composition and dependence among observations.

\subsection{Cross-source integration of trust dimensions and contextual boundaries}
\label{subsec:integration}

Across the three sources, convergence was strongest around capability and epistemic scrutiny, although each source captured these issues at a different level (Table~\ref{tab:joint-display}). The survey identified instrumental and transparency-grounded trust as individual-level dimensions, interviews showed how capability, verification, and evidentiary requirements entered situated trust judgments, and Weibo discourse located related concerns within socially visible discussions of productivity, credibility, education, misinformation, and governance. Relational responsiveness also appeared across sources, forming a clear survey dimension and appearing in interview accounts of emotional and socially appropriate interaction, while remaining comparatively less prominent in public discourse.

Cross-source comparison also clarified the boundary of the measure. Privacy, accountability, copyright, misinformation, and task-specific risks emerged as broader concerns surrounding trust and reliance. These concerns fall outside the scope of the retained instrument but remain important for understanding the contexts in which trust judgments are formed and acted upon. The resulting scale therefore captures three related bases of generalized GenAI trust---instrumental, relational, and transparency-grounded trust---while the qualitative and computational evidence situates these judgments within the broader social ecology of GenAI use.

\begin{table}[H]
\centering
\caption{Joint display of trust dimensions and contextual evidence across data sources}
\label{tab:joint-display}
\scriptsize
\setlength{\tabcolsep}{4pt}
\begin{tabularx}{\textwidth}{@{}
  >{\raggedright\arraybackslash}p{2.0cm}
  >{\raggedright\arraybackslash}p{2.1cm}
  >{\raggedright\arraybackslash}p{2.8cm}
  >{\raggedright\arraybackslash}p{2.7cm}
  Y@{}}
\toprule
Content area & Survey & Interview & Public-discourse & Cross-source interpretation \\
\midrule

Instrumental capability
& INS: three competence items
& Capability and accuracy shaped whether AI was considered dependable across tasks
& Work and productivity was the largest topic; education and other application contexts were also prominent
& Convergence on system capability, with interviews and discourse specifying where performance matters \\

\addlinespace
Transparency and epistemic scrutiny
& TRA: explanations of system operation, outputs, and information sources
& Accuracy, verifiability, labeling, source disclosure, and human checking shaped confidence and reliance
& Credibility and transparency formed a distinct topic, while related concerns also appeared in misinformation, education, and governance
& Convergence on epistemic scrutiny; interview evidence links the factor to verification practices, while discourse evidence identifies related issue contexts \\

\addlinespace
Relational responsiveness
& REL: benevolence, emotional understanding, social competence, care, and warmth
& Emotional support and socially appropriate interaction appeared in selected task contexts
& Emotional interaction and anthropomorphism formed a smaller but comparatively positive topic
& Partial convergence; the dimension was clearer as a generalized individual-level judgment than as a dominant public issue \\

\addlinespace
Privacy and institutional accountability
& Outside the retained trust dimensions
& Privacy, authorization, platform responsibility, and accountability appeared as risks or conditions of trust
& Privacy, governance, and fairness formed salient and often negatively evaluated issue domains
& Boundary evidence: primarily trust-relevant conditions, perceived risks, and governance concerns rather than reflective trust dimensions \\

\addlinespace
Task and social consequences
& Outside the retained trust dimensions
& Trust thresholds varied across news, education, work, content creation, and other uses
& Labor, misinformation, copyright, education, and long-term consequences structured public discussion
& Extension: contexts in which generalized trust judgments acquire practical significance \\

\bottomrule
\end{tabularx}
\end{table}

\section{Discussion}
\label{sec:discussion}

\subsection{Theoretical contribution}

This study extends technology-trust research by conceptualizing public trust in GenAI as an integrated multidimensional judgment organized around three related but distinguishable grounds for warranted confidence. Instrumental trust reflects expectations about what the system can do, retaining the competence-oriented logic emphasized in established technology-trust models, whereas relational trust reflects expectations about how the system interacts with users \citep{McKnight2011SpecificTechnology,lankton2015Technology}. Transparency-grounded trust captures whether explanations, traceable sources, and opportunities for verification provide users with epistemic resources for evaluating system operation and outputs. This dimension is particularly salient for generative systems, whose conversational interfaces and probabilistic outputs make both relational responsiveness and the inspectability of generated content consequential to trust. 

The empirical relationships among these dimensions are consistent with this interpretation. Their moderate-to-strong correlations indicate a shared orientation toward confidence in GenAI, while their distinct loading patterns and HTMT estimates support their empirical distinction. The comparatively stronger association between instrumental and transparency-grounded trust is also theoretically meaningful: judgments of system capability may be difficult to separate from the explanations and evidentiary resources that make performance inspectable. The resulting structure therefore captures explanations and verifiability as epistemic conditions that justify user confidence, separate from abstract perceptions of system openness.

The cross-source evidence also helps clarify the scope of the resulting measure. 
Privacy, accountability, copyright, misinformation, and labor displacement emerged as broader concerns surrounding GenAI trust and reliance in interviews. These concerns fall outside the scope of the retained measure and help explain the social contexts in which trust judgments are made. This interpretation is consistent with established distinctions among trust, perceived trustworthiness, attitudes, and reliance \citep{McKnight2011SpecificTechnology,McGrath2025Measuring}.

The computational findings show how the issues surrounding GenAI trust changed over time. Consistent with research showing heterogeneous public responses to AI \citep{KelleyEtAl2021EightCountries,qi2024Excitements,zhou2026AI}, Weibo discourse combined productivity opportunities with concerns about misinformation, privacy, fairness, and governance. More importantly, temporal change was much clearer in issue composition than in standardized stance within comparable issue contexts. The late-August deepfake surge illustrates how a salient event can rapidly reshape the distribution of AI concerns, whereas standardized stance showed no corresponding systematic temporal shift within comparable observed discourse contexts. These findings therefore contextualize the multidimensional trust judgments identified in the survey within changing issue agendas, which shape the vulnerabilities and safeguards that become salient at a given moment.

\subsection{Methodological contribution}

Public salience and psychometric dimensionality need not coincide. A concern that dominates public discourse may fall outside the scope of an individual-level trust measure, while a distinct trust dimension may receive relatively little public attention. Integrating qualitative, psychometric, and computational evidence helps identify these differences and clarify the scope of measurement. The joint display illustrates how socially visible concerns, situated experiences, and survey responses can inform the interpretation of a construct.

The computational component also contributes to longitudinal discourse analysis. Building on studies that use large-scale text data to trace issue salience, evaluation, and temporal change \citep{qi2024Excitements,zhou2026AI}, the analysis examined whether temporal patterns were sensitive to retrieval composition, concentrated events, and dependence among posts. Dependence-aware and event-window analyses showed clear change in overall issue composition, while several individual topic trajectories were sensitive to cross-day dependence and late-period event concentration. By contrast, source- and topic-standardized stance showed no systematic temporal shift. This distinction is particularly important for keyword-retrieved social-media corpora, where event-driven changes in corpus composition can otherwise be mistaken for broader changes in evaluative stance.

\subsection{Practical implications}

The findings suggest that AI governance and design should support calibrated trust. The association between instrumental and transparency-grounded trust highlights the importance of epistemic resources in users' evaluations of GenAI. Explanations, traceable sources, and accessible fact-checking tools can help users assess system outputs and make informed decisions about reliance, particularly in consequential settings.

Relational design requires a complementary safeguard. Warmth, empathy, and socially appropriate communication may improve interaction, but they should not substitute for evidence of factual reliability or institutional accountability. In advisory, educational, or companionship settings, socially engaging interfaces should therefore be paired with clear boundaries, source information, and routes to human oversight.

The temporal findings also point toward event-responsive governance. The late-August deepfake surge illustrates how high-profile incidents can rapidly reshape which AI-related concerns dominate visible platform attention. AI literacy initiatives may likewise be more effective when organized around concrete tasks and periods in which particular risks or verification needs become especially salient.

\subsection{Limitations and future research}

The Weibo corpus represents visible platform discourse, with its composition influenced by retrieval terms, platform participation, and event-driven attention. Classification performance varied across topics and stance categories, and the two-stage screening procedure may have introduced uneven coverage of relevant discourse. Future research could combine multiple platforms and event-centered designs to examine changes in public discourse.

The psychometric evidence is based on a cross-sectional online-panel sample of GenAI-aware adults in China. Its generalizability requires further investigation. The interviews were conducted in a single setting and capture a limited range of situated experiences. Instrumental and transparency-grounded trust showed empirical overlap, and the conditional wording of the transparency-grounded items may capture a different response process from the direct evaluative statements used in the other subscales. Future research should examine alternative item formulations, measurement invariance, temporal stability, and associations with observed verification and reliance behavior.

\section{Conclusion}
\label{sec:conclusion}

This study develops a ten-item measure of public trust in GenAI in China, encompassing instrumental, relational, and transparency-grounded dimensions. Interview and computational evidence situate these dimensions within broader concerns about verification, risk, and institutional responsibility. The Weibo analysis identifies changes in issue composition between 2023 and 2024, with no evidence of a systematic shift in evaluative stance after accounting for retrieval source and topic composition. By integrating evidence across these analytical levels, the study provides a foundation for examining how people evaluate and rely on GenAI across social contexts.

\section*{Declarations}

\paragraph{Supplementary data}
The Supplementary Material contains Appendices~A--G. Supplementary Table~S1 is supplied as a separate CSV with the full set of temporal sensitivity estimates for the 12 topics.

\paragraph{Data availability}
Data will be made available on request.

\paragraph{Declaration of competing interests}
The authors declare that they have no known competing financial interests or personal relationships that could have appeared to influence the work reported in this paper.

\paragraph{CRediT authorship contribution statement}
Zhiyuan Lai: Conceptualization, Methodology, Software, Formal analysis, Investigation, Data curation, Visualization, Writing--original draft, Project administration.
Wanxin Ran: Methodology, Data curation. 
Pu Yan: Conceptualization, Methodology, Resources, Supervision, Writing--review and editing.
Zhouhan Chen: Software, Investigation, Data curation.

\bibliographystyle{elsarticle-harv}
\bibliography{genai-trust-ti}

\end{document}